\documentclass[11pt]{article}
\usepackage{color}
\usepackage{amssymb}
\usepackage{amsmath}
\usepackage{epsfig}

\definecolor{pink}{rgb}{1,.4,.7}
\definecolor{magenta}{rgb}{1,0,1}
\definecolor{violet}{rgb}{.9,.25,.6}
\definecolor{darkolivegreen3}{rgb}{.6,.8,.35}
\definecolor{maroon3}{rgb}{.8,.26,.56}
\definecolor{mediumorchid}{rgb}{.73,.33,.83}
\definecolor{mediumorchid1}{rgb}{1.,.33,.63}
\definecolor{darkgreen}{rgb}{0.1,.6,.13}
\definecolor{lightyellow}{rgb}{1.,1.,.82}
\definecolor{turquoise}{rgb}{.35,.80,.71}
\definecolor{coral}{rgb}{1.,.6,.21}
\definecolor{orangered}{rgb}{1.,.5,0.}
\definecolor{orange}{rgb}{1.,.65,.1}
\definecolor{blue1}{rgb}{.48,.53,1.}
\definecolor{gold}{rgb}{1.,.85,0.}
\definecolor{darkviolet}{rgb}{.54,.04,.84}

\def\Journal#1#2#3#4{{#1} {\bf #2}, #3 #4}

\def\etal{{\it et al.}}
\def\CQG{\em Class.Quant.Grav.}

\def\GRG{\em Gen. Rel. Grav}

\def\IMD{{\em Int. J. Mod. Phys.} D}

\def\JCA{\em J. Cosmol. Astrop. Phys.}

\def\PRD{{\em Phys. Rev.} D}
\def\PRL{\em Phys. Rev. Lett.}

\def\be{\begin{equation}}
\def\ee{\end{equation}}
\def\bea{\begin{eqnarray}}
\def\eea{\end{eqnarray}}
\begin{document}
\begin{center}
\Large \bf {A note about the back-reaction of inhomogeneities on the expansion 
of the Universe}\\
\end{center}

\begin{center}
{\it Houri Ziaeepour\\
Mullard Space Science Laboratory\\
Holmbury St. Mary, Dorking, RH5 6NT Surrey, UK\\
{Email: {\tt hz@mssl.ucl.ac.uk}}
}
\end{center}

%\medskip
%\begin {tabular}{p{12cm}p{3cm}}
% & \bf \it {To Memory of Changoul and her Love}
%\end {tabular}
%\medskip

\begin {abstract}
In this short note we summarize the arguments against a significant 
back-reaction of inhomogeneities on the acceleration of the Universe. 
We also present a quick way to access the importance of back-reaction using 
the Fourier space presentation of inhomogeneities and properties of their 
power spectrum.
\end {abstract}

\section* {}
We first consider the suggestion of dark energy as the effect of second and 
higher order terms in a perturbative treatment of an inhomogeneous 
universe~\cite{brkolb0,brkolb1,brkolb2,brkolb3}. The essential point of 
reasoning in these works is the first (or second) order expansion of the 
metric $g_{\mu\nu}$, Einstein tensor $G_{\mu\nu}$, and energy-momentum tensor 
$T_{\mu\nu}$ in synchronous gauge:
\bea
ds^2 &=& a^2(\tau)[d\tau^2 - \gamma_{ij} (\tau, \mathbf{x}) dx^idx^j] \label{metric} \\
\gamma_{ij} &=& (1-2\psi^{(1)})\delta_{ij} + D_{ij}\chi^{(1)} + 
\partial_i\chi_j^{(1)} + \partial_j\chi_i^{(1)} + \chi_{ij}^{(1)} \label{gammaij} \\
G_{\mu\nu}^{(0)} + G_{\mu\nu}^{(1)} &=& 8\pi G (T_{\mu\nu}^{(0)} + T_{\mu\nu}^{(1)}) \label{einsteineq}
\eea
where $\psi$, $\chi$, $\chi^i$, and $\chi^{ij}$ are respectively scalars, 
vector, and tensor fields characterizing an inhomogeneous FLRW universe up to 
the first order. Then up on averaging:
\be
G_{\mu\nu}^{(0)} = 8\pi G \langle T_{\mu\nu} \rangle - \langle G^{(1)}_{\mu\nu} \rangle \label{einsteineqp}
\ee
Note that $T_{\mu\nu}$ in (\ref{einsteineqp}) is the exact 
energy-momentum tensor. Therefore the term $\langle G^{(1)}_{\mu\nu} \rangle$ 
actually includes all the inhomogeneities not just the linear order. 
The claim is that this term can play the role of dark energy and imitate 
an accelerating Universe. The error in such a claim is 
very obvious. Any perturbative expansion contains an expansion scale 
$0 \leqslant \epsilon \leqslant 1$, which if it approaches zero, the expanded 
quantity approaches the zero order. The expansion makes sense only if the 
higher order terms are at most $\sim{\mathcal O}(\epsilon)$\footnote{A number 
calculation of perturbations up to second order show that it can be more 
important than the linear term~\cite{secperturb}. Therefore 
${\mathcal O} \epsilon$ should be interpreted as the dominant term}. 
Thus, assuming that at cosmological scales of interest inhomogeneities are 
small, the dominant correction are ${\mathcal O}(\epsilon) \rightarrow 0$, 
much smaller than the zero-order quantities. This means that the claimed 
contribution of inhomogeneities can not produce a dark energy density 
$\sim 2.5$ times larger than the underlaying matter, otherwise a perturbative 
treatment of density perturbations does not make sens. Moreover, by definition:
\be
\langle T_{\mu\nu} \rangle = T_{\mu\nu}^{(0)}
\ee
meaning that:
\be
\langle G^{(1)}_{\mu\nu} \rangle = 0
\ee
or if $T_{\mu\nu}^{(0)}$ is defined or observationally determined such that 
$\langle T^{(1)}_{\mu\nu} \rangle \neq 0$, then as the terms with the same 
order in the two sides of (\ref{einsteineq}) should cancel each other, the 
effect of averaged first and higher order Einstein tensor is canceled by the 
effect of averaged matter inhomogeneities. Considering very detailed 
calculations in~\cite{brkolb0,{brsecond}}, the argument presented here seems 
simplistic. Nonetheless, it is based on the rules and concepts which must be 
respected by any perturbative expansion irrespective of the underlaying 
physics. A number of works~\cite{responpap} have discussed the 
inhomogeneities effect by detailed perturbative calculation up to the second 
order and in particular they show that the claimed effect of super-horizon 
perturbations~\cite{brkolb1,brkolb2,brkolb3} is small. See below for another 
demonstration of this effect.

There is one issue that has not been considered in any of these works. In a 
perturbative expansion although each order can be much smaller than lower 
orders, the sum of all the terms can diverge. This a well known fact for 
instance in the perturbative expansion of quantum electrodynamics (QED) 
effective Lagrangian~\cite{qedborel}. Can a divergence arise in the 
perturbative expansion of the cosmological inhomogeneities ? The answer is 
no. The reason is the observational evidence for a scale-less power spectrum 
of anisotropies. Consequently, anisotropies with significant variations - 
large $k$ in Fourier space - are not dominant. The back-reaction due to mixing 
of IR and UV scales is either a pure gauge~\cite{secperturb} or can be removed 
be renormalization~\cite{responpap}. The scale-less spectrum also means that 
they cover a very small space in the volume of the Universe. Therefore their 
effect on the global expansion is effectively negligible, in another word 
the perturbative expansion in (\ref{einsteineq}) does not diverge. See below 
for a quantitative demonstration of this claim. In an observational language, 
at cosmological distances light propagates roughly on a straight line and its 
average deflection is very small.

The issue of averaging is presented by Buchert~\cite{avbuchert} in more 
mathematically rigorous way. Nonetheless, the claim of implication of 
averaging in dark energy paradigm can not be true. Essentially, the idea is 
that cosmological measurements are always performed in a finite volume of a 
space-like foliation of the spacetime and we can extend this volume at most 
to the observer's horizon. Moreover, due to the expansion of the Universe the 
volume enclosing a given set of particles evolves. Consider the following 
foliation, called Lagrangian coordinates in a synchronous gauge:
\be
ds^2 = -dt^2 + g_{ij} dX^i dX^j \label{metriclagrangian}
\ee
For any scalar quantity $\psi (X,t)$ averaging in a constant-time volume $V_D$ 
is defined as:
\be
\langle \psi \rangle_D \equiv \frac{1}{V_D} \int_{V_D} d^3 X J \psi, 
\quad \quad J = \sqrt{|\det g_{ij}|}, \quad \quad V_D \equiv 
\int_{V_D} d^3 X J \label{aver}
\ee
Here only a dust matter is considered but the formulation can be extended to 
other type of fluids~\cite{avbuchert}. The averaged expansion factor $a_D$ 
is defined as:
\be
a_D (t) \equiv \biggl (\frac{\dot{V}_D}{V_D} \biggr )^{1/3}
\ee
Few other definitions and relations:
\bea
&& K_{ij} \equiv \frac{1}{3} \theta \delta_{ij} + \sigma_{ij}, \quad \quad 
\sigma^2 \equiv \sigma^{ij}\sigma_{ij} \label {kdef}\\
&& \dot J = \theta J, \quad \quad \langle \theta \rangle_D = \frac{\dot{V}_D}
{V_D} = \frac{3\dot{a}_D}{a_D} \label {jav} \\
&& \langle \partial_t \psi \rangle_D - \partial_t \langle \psi \rangle_D = 
\langle \theta \psi \rangle_D - \langle \theta \rangle_D \langle \psi \rangle_D
\label {timevar}
\eea
where $K_{ij}$ is the extrinsic curvature. In the same way the volume 
averaged analogue of $\Omega_m$ and $\Omega_\Lambda$ in homogeneous FLRW 
cosmology can be defined using average quantities.

It can be shown that averaging adds a term $Q_D$ called back-reaction to the 
Einstein and mass conservation equations written for the finite volume 
averaged quantities:
\bea
&& 3\biggl (\frac{\dot{a}_D}{a_D} \biggr )^2 - 8\pi G \langle \rho \rangle_D + 
\frac{1}{2} \langle R \rangle_D - \Lambda = -\frac{Q_D}{2} \label{hubbleq}\\
&& \frac{3\ddot{a}_D}{a_D} + 4 \pi G \langle \rho \rangle_D - \Lambda = Q_D 
\label {enerconv} \\
&& Q_D \equiv \frac {2}{3} \langle (\theta - \langle \theta \rangle_D)^2 
\rangle_D - 2 \langle \sigma^2 \rangle_D \label {enerconv}
\eea
Form (\ref{hubbleq}) we can see that if $Q_D < 0$, it can play the role of a 
{\it new matter}. In particular, if it does not change quickly with 
time/redshift, it can play the role of a dark energy. However, there are 
various arguments against such a possibility:
\begin{itemize}
\item The first and most evident issue is that the Lagrangian averaging -  
following the same set of particles and their evolution - 
considered here does not correspond to the way we observe cosmological 
objects. Large scale observations are based on an Eulerian concept, i.e. 
sampling different part of the cosmic fluid at different time. Assuming that 
our observations at a given redshift is complete, only the matter outside 
the observer's horizon can have an unaccounted for effect on the global 
expansion of the visible Universe. This issue is 
related to the claim of super-horizon anisotropies effect on the acceleration 
of the Universe\cite{brkolb1,brkolb2,brkolb3}. As mentioned above a number of 
works~\cite{responpap} have shown that this effect is negligible (See also 
below for another demonstration). Moreover, 
$Q_D$ is strongly gauge dependent and there are examples showing that it can 
be negative without having a real accelerating universe~\cite{brward}.
\item Forgetting the problem with observations, we try to estimate $Q_D$ 
in a universe with small inhomogeneities like our Universe. The extrinsic 
curvature is defined as:
\be
K_{ij} = - h^\alpha_i h^\beta_j u_{\alpha;\beta}, \quad \quad 
h_{\alpha\beta} = g_{\alpha\beta} + u_\alpha u_\beta \label{extrinsic}
\ee
Using (\ref{extrinsic}) and (\ref{kdef}) it is easy to see that in a 
perturbative expansion of the metric $g_{ij}$ and matter density $\rho$, 
scalar fields $\theta^2$ and $\sigma^2$ are of second order and therefore 
if $V_D$ is enough large such that the linear regime approximation can be 
applied inside this volume, $Q_D$ is negligible. Even if the second order 
term is more important, the total effect of all the terms must be at most of 
first order and negligible with respect to the dominant homogeneous effect. 
\item Another way of seeing that the effect of finite averaging volume is 
negligible if inhomogeneities are perturbative, is through (\ref{timevar}).
It is evident that the difference between two terms in the right hand side of 
(\ref{timevar}) is of second order and therefore negligible in linear 
regime. Note that averaging is in real space and the confusion from 
mixing of large and small scale does not arise. Moreover, the second order of 
the right hand side corresponds to the second order of l.h.s. because the 
linear orders in each side cancel out exactly. In addition, in cosmological 
context most of the physically interesting scalar 
fields such as the matter density $\rho$ are random fields. In this case if 
the volume $V_D$ is enough large, the  ergodicity of the random field makes 
the left side of (\ref{timevar}) to vanish. This argument is valid without 
assuming a perturbative approximation.
\item Because $Q_D$ depends on the shear $\sigma$ and on the curvature 
$\theta \propto -K$ where $K$ is the trace of the extrinsic curvature, a large 
amount of curvature, comparable to the horizon size, is needed to explain 
observations~\cite{brbuchert0,brbuchert1}. The observed curvature of the 
Universe at all scales - from local Universe up to CMB time - is very small. 
On the other hand models suggested in~\cite{brbuchert0,brbuchert1} have all 
a curvature which at most is constant i.e. decays with time as $a_D^{-2}$ up 
to as fast as $a_D^{-6}$. We copy here the summary conclusion of 
Ref.~\cite{brbuchert1}(page 24) on this subject:

{\it Dark energy cannot be routed back to inhomogeneities on large scales in 
Newtonian and quasi-Newtonian models, but a careful re-interpretation of 
cosmological parameters will have nevertheless to be envisaged.}

The solution which has been then suggested to make the back-reaction effect 
sufficiently important and at the same time consistent with observations, is 
a static out-of-equilibrium universe with 
large curvature~\cite{brbuchert0,brbuchert1}. It is not necessary to mention 
that present data as well as our knowledge from the microphysics of the 
Universe does not support such a solution for the origin of dark energy.
\end{itemize}

Another way of seeing the effect of both super-horizon and sub-horizon 
perturbations is by using the finite volume averaging eq.(\ref{aver}). 
It is easy to see that the volume average of a scalar field $\psi$ is:
\be
\langle \psi \rangle_D = \int d^3 k \psi' (k) \prod_{i=1}^3 sinc (k_i x_D^i), 
\quad \quad \psi'(k = 0) = \langle \psi \rangle_\infty \label{volfourier}
\ee
where $\psi' (x) = J\psi$ and $x_D^i$ is a characteristic size scale of the 
volume $V_D$ in $x^i$ direction - for a cube parallel to the axes, it is the 
length of the edge parallel to axis $i$. When inhomogeneities spectrum is 
scale-less, from properties of $sinc$ function we can conclude that the 
contribution of $k \gg 1/x_D$ i.e. inhomogeneities at scales much smaller 
than $X_D$ are negligible. The contribution of $0 < k < 1/X_D$ modes is 
proportional to $1/k$. Therefore, in the case of a scale-less spectrum where 
statistically averaged value of $\psi' (k)$ is mode independent, the 
integral over these modes after renormalization of IR divergence is a 
sub-dominant logarithmic term $\propto \log x_D^{-1}$ which in an inflationary 
universe is very small. This confirms the results of ref~\cite{responpap} and 
shows that the right hand side of (\ref{timevar}) is very close to zero. 
Therefore, the assumption of commutation between time and space averaging is a 
good approximation in a close to scale-less universe. The difference between 
finite and infinite volume decreases with the expansion of the Universe in 
contrast to the dark energy which become dominant at late times. Note that 
the argument given here is only based on the statistical properties of 
inhomogeneities and a perturbative behaviours has not been assumed. In 
summary inhomogeneities, sub-horizon or super-horizon, can not explain the 
observed dark energy component of the Universe.


\begin{thebibliography}{99}
\bibitem {brkolb0} Kolb E.W., \etal, \Journal{\PRD}{71}{2005}{023524}, hep-ph/0409038.
\bibitem {brkolb1} Baraausse E, \etal, \Journal{\PRD}{71} (2005){063537}, astro-ph/0501152.
\bibitem {brkolb2} Kolb E.W., \etal, hep-th/0503117.
\bibitem {brkolb3} Kolb E.W., \etal, arXiv:0901.4566.
\bibitem {brsecond} Losic B., Unruh W.G., \Journal{\PRD}{72}{2005}{123510}, gr-qc/0510078.
\bibitem {secperturb} Mukhanov V.M., Abramo L.R., Brandenberger R.H., \Journal{\PRL}{78}{1997}{1624}, gr-qc/9609026,\\ 
Abramo L.R., Brandenberger R.H., Mukhanov V.M. \Journal{\PRD}{56}{1997}{3248}, gr-qc/9704037, \\ 
Unruh W, astro-ph/9802323,\\
Geshnizjani G., Brandenberger R., \Journal{\PRD}{66}{2002}{123507}, gr-qc/0204074,\\
Geshnizjani G., Brandenberger R., \Journal{\JCA}{0504}{2005}{006}, hep-th/0310265,\\
Losic B., Unruh W.G., \Journal{\PRD}{74}{2006}{023511}, gr-qc/0604122.
\bibitem {responpap} Gishnizjani G., \etal, \Journal{\PRD}{72}{2005}{023517}, astro-ph/0503553,\\
Flanagan \'E.\'E, \Journal{\PRD}{71}{2005}{103521}, hep-th/0503202,\\
Zeng D. \& Gao Y., hep-th/0503154,\\
Hirata C.M. \& Seljak U, \Journal{\PRD}{72}{2005}{083501}, astro-ph/0503582,\\
Larena J, \etal, \Journal{\PRD}{79}{2009}{083011}, arXiv:0808.1161.
\bibitem {qedborel} Dunne G.V. and Hall T.M., \Journal{\PRD}{60}{1999}{065002},  hep-ph/990206.
\bibitem {avbuchert} Buchert T., \Journal{\GRG}{32}{2000}{105}, gr-qc/9906015,\\
Buchert T., \Journal{\GRG}{33}{2001}{1381}, gr-qc/0102049, \\
Buchert T., \Journal{\CQG}{19}{2002}{6109}, gr-qc/0210037, \\
Buchert T., Carfora M., \Journal{\PRL}{90}{2003}{031101}, gr-qc/0210045, \\
Buchert T., Proc. 12th JGRG conference, Tokyo 2002, Shibata M. \etal (eds.) (2003), 157-161, astro-ph/0312621.
\bibitem {brward} Ishibashi A. \& Ward R.M., \Journal{\CQG}{23}{2006}{235}, gr-qc/0509108.
\bibitem {brbuchert0} Buchert T., \Journal{\CQG}{22}{2005}{L113}, gr-qc/0507028
\bibitem {brbuchert1} Rasanen S., \Journal{\IMD}{15}{2006}{2141}, astro-ph/0605632,\\
Buchert T., \Journal{\GRG}{40}{2008}{467}, arXiv:0707.2153.
\end{thebibliography}
\end{document}